# Manganese oxide nano flakes onto simultaneously activated defective CNTs for the creation of CNTs-graphene supported oxygen reduction composite catalysts applied in air fuel cell

**Ling Sun** [a, c] [*] and Danxian Liu [b, c]

*a. Beijing Guyue New Materials Research Institute, Beijing University of Technology, 100 Pingleyuan, Chaoyang District, Beijing 100124, China.*

*b. Material and Industrial Technology Research Institute Beijing, No.166, Est Street, Liupingzhuang, Doudian Zhen, Fangshan District, Beijing, 102402, China.*

*c. Advanced (Energy) Materials Joint Laboratory, Beijing University of Technology, 100 Pingleyuan, Chaoyang District, Beijing 100124, China.*

## ABSTRACT

Rational combination holds the key for non-precious metal oxides and economic carbon materials to the creation of high performance air electrode materials. Carbon nanotubes (CNTs)-supported manganese oxides composite catalysts (CMnCs) were obtained by reacting commercial CNTs with $MnO_4^-$ in diluted $H_2SO_4$ with cancelling conventional hydrothermal processing or adding surfantant templates. CMnCs were then assembled into commercial air cathodes equipped in metal air fuel cells. CNTs demonstrated structurally much defective. Outer CNT walls were cracked into graphene nano pieces yet remaining partially attached to inner carbon tubes during synthesis. This exposure of graphene planes exhibited additional oxygen reduction reactivity. Manganese oxide flakes identified as γ-MnOOH were homogeneously grown on surfaces of CNTs at nano sizes. Its deposition density was closely related to the additive amount of $KMnO_4$ with regard to CNTs. The ORR activities of sole CMnCs exceeded both raw CNTs and competent market catalysts under identical conditions, and with current production technics the CMnCs-ingredented air fuel cells were capable of outputting ~15% more power at 100 mA/cm$^2$.

[*] Corresponding author. E-mail: sunling@bjut.edu.cn; captainsun@ees.hokudai.ac.jp

1. **Introduction**

Outdoors sports are becoming more prevalent in developing countries than ever, in the meantime, environmental emergencies are also frequently reported accompanied with their giant economic achievements. We are aware of that to face these scenarios, a growing need of durable energy-releasing devices becomes more urgent than before, especially those capable of tens of hours' electric supporting for various electronics. Metal air fuel cells (AFC, metal: Al, Mg, Zn, Fe, etc.) are important power sources because of their remarkable high theoretical energy density and low cost [1,2]. An AFC in general comprises a metal plate anode and an oxygen reduction reaction (ORR)-oriented catalytic cathode (including gas diffusion electrodes). Its variety is normally determined by the exhaust anodes, e.g., water-borne type for zinc, iron, magnesium fuel cells, organic type for lithium and either for aluminium. All these years, water-borne air fuel cells AFC are one important topic across academic research and industrial field. The exclusive use of platinum-group metal-free ORR catalysts become ubiquitous and transition metal oxides, like manganese oxide, have been the typical one of those cost-effective ORR catalysts [1,3-8]. The technical maturity and economical efficiency had greatly motivated abundant researches in and just recently, we successfully industrialized one technology to manufacture water-type oxide film electrodes capable of being equipped into commercial magnesium AFCs (http://www.mitbj.org/aspcms/product/2016-11-11/473.html).

As mentioned, such electrodes are dedicated to emergent long-time illuminating use at a relative low output power. With expectation for higher-power vehicle systems as the development target, a further elevating on this index remains challenging. The first of all steps comes to rational design and manufacture to further decrease the electrical resistance and compactness. To this end, current solutions include: (1) catalysts are made nano/micro sized to get more accessible surfaces, and (2) active components are physically mixed with or chemically bonded to highly electrically conductive substrates which construct electron-delivering pathways, for example, low-dimensional high-aspect-ratio nano materials. While in production lines, mechanical mixing is normally a cost-effective preferential choice, and micro-sized or even larger, yet cheap conductive additives are normally included, such as activated carbon, conductive acetylene black (AB), etc. AB are string-like quasi-one-dimensional structure [9], and with this longitudinal advantage, is indeed helpful to improve the inner resistance, and subsequent catalytic electrodes displayed better ORR activity at the initial stage. But it's worthy of noticing, (1) at high output power (in lab, 100 w/unit) the temperature of the working cell easily reached over 60℃ and fluctuated up even much higher, and (2) in a long-time discharge, these electrodes were seen with volumetric swelling when frequently replacing anodes and electrolytes. These resulted in further spacing to the physical connection between the catalysts and those additives. To avoid the consequent conductivity decay chemical bonding was one substitutional method [9, 10] and in this case we used one-dimensional commercially available multi-walled carbon nanotubes (CNTs).

Comparably to AB, CNTs are in nature more electrically conductive, and grow more available in variable lengths, e.g., tens of, and hundreds of micrometres, even the record-breaking half meter [11, 12]. They have been industrialized and become commercially economic, good candidates as micro-sized current collectors/supporters for electrode materials selection. In fact, they have been used as conductive additives to make lithium-ion batteries and capacitors in countries like China and Japan, or as strengthening additives in metal alloy via power metallurgy method [13], etc. We more cared about that, some previous research had reported CNTs themselves ORR-active, especially those with the tube structures partially unzipped and semi-attached graphene structures exposure to outside [14, 15]. The surface structure and conductance features of these modified CNTs conclusively played important roles on ORR performances.

Thus, with graphene involvement, the combination of CNTs and conventional oxides would potentially innovate conventional ways to make ORR effective composite catalysts. How to evenly grow the oxide nanostructures onto CNTs surfaces became the key step. To date, several common oxidization-reduction methods have been frequently reported using $KMnO_4$ as manganese precursors, some of which resembled the Hummers' method (used to prepare GO from flake graphite [14, 16]). It's known oxidative in all pH, especially strong in acidic. Reacting with CNTs began from their outer-wall deformation mainly resulting from the oxidation of the existing defects, and simultaneously involved another two-step processes: defects generating, such as -OH and –C=O, by attacking the graphene structure through electrophilic reactions and defects consuming through breaking up the neighbouring graphite structures around the above defects [17]. During such, high-valence Mn(VII) were reduced to Mn(II), Mn(III) or Mn(IV) in oxide or salt forms, depending on real constraint reactive conditions. We compared and noticed that for the purpose of catalyst oxides, the above methods need extra measures, such as the addition of surfactants [18], high temperature treatments [10, 14, 18, 19] or pre-treated/specific precursors [10, 20, 21].

To further economize this process, we directly adopted in-market raw CNTs, which from Raman microscopy demonstrated rich structural defects, and chose diluted sulfuric acid aiming to assisting to (1) oxidize/etch the CNTs outer-wall structures, resembling to the commonly-seen "activation" of carbon, and (2) clear off the amorphous carbon deposits on tube surfaces that were originally brought in from manufacturing, thus avoiding using concentrated acid and inducing unexpected hazards. Consequently, the synthesis turned out with only mixing and stirring at room temperature, being rather mild, very similar to previous report with AB as carbon supporter [9] in spite of the acid. As a result, we found that manganese oxide deposits were evenly distributed on the CNT (graphene) supporter forming a core-shell hierarchical structure. These oxides were flaky in shape and nano thick, with a typical γ-MnOOH structure. The formation of semi-attached graphene nano pieces demonstrated to exist on the carbon tubes as implied from the electrochemical characterization. Eventually manganese oxide-CNTs composite catalysts were obtained displaying in the expected composite structure, and denoted as CMnCs throughout this research. After integrated characterizations, CMnCs catalysts were in factory made into air electrodes and assembled into AFCs.

## 2. Experimental

### 2.1 Chemicals and reagents

CNTs with a product code NC7000, were commercially available. Manganese oxide (denoted as the Ref) was acquired from MITBJ and used as a reference counterpart in this research. Reagents such as potassium permanganate, sulfuric acid, were of high-purity grade and were used as received without any further purification.

### 2.2 Synthesis of CMnCs

Prior to use, CNTs were ultrasonically washed in ~60 ℃ deionized water, filtrated and dried. After that, CNTs (0.04 g/ml) were suspended into sulfuric acid (0.2 mol/l, 160 ml) and sonicated for 10 minutes at the frequency of 49 KHz. Potassium permanganate was then added at designed series of concentrations of 0.4, 0.2, 0.1 mol/l and dissolved through a vigorous stirring. The mole ratio of $KMnO_4$ versus CNTs (K/C) rose from 3.2%, 6.4% through 12.8%. All beakers were kept at room temperature (~20 ℃) for ~4 hours until inside solutions being completely colourless. The resultants were collected by filtration, then fully washed with deionized water and vacuum dried in 80 °C overnight. Then the CMnCs powders were obtained. All these catalysts were assembled to air electrodes and AFCs in pilot factory for subsequent property investigations.

### 2.3 Characterizations and measurements

Morphological characterization and elemental mapping were executed by using a field emitter scanning transmission electronic microscope (STEM, SU9000, Hitachi, Japan) at the accelerating voltage 5kV.

For electrochemical measurement, we used a three-electrode system. The working electrode was prepared as following steps: each sample of 5 mg was ultrasonically mixed with 0.9 ml ethanol and 0.1 ml 5 % commercial Nafion liquid to prepare a suspended liquid; then several drops with a total volume of 20 μL were respectively dripped onto surfaces (~ 1$cm^2$) of glassy carbon (GC) electrodes and these electrodes were put under an infrared ray lamp until the solvents were vaporized off with black tense films left. After cooled to room temperature, the working electrodes were respectively immersed in alkaline solution (0.1 M KOH) with a platinum foil (20 mm×20mm) as the counter electrode and a Ag/AgCl standard electrode (R0305, Tianjin Ida, China) as the reference. A conventional electrochemical workstation (Chi66e, CH

Instruments Inc.) was connected with them. Measurements including cyclic voltammetry (CV) with a scan rate at 0.05 V/s in a scan range of -1~ 0.2 V vs. Ag/AgCl at room temperature ~20 ℃ were executed.

To prepare CMnCs cell, the Ref was substituted in the product ingredients and the state-of-the-art route of MITBJ was applied. Detailed description of making electrode has been patented (No. CN205645985U). Thus electrodes were able to be in-factory made and tested with a polished commercial zinc plates (25mm × 35mm) as counter electrode, in 6 M KOH solution using a CT-3008 system (Neware Technology Co. Ltd, Guangdong, China). Note: areal content of CMnCs ~0.19 mg/cm$^2$, exposed area to electrolyte for air electrode ~10 cm$^2$, distance between electrodes ~1cm. MITBJ has leadingly developed high-performance metal-air electrodes/batteries with catalyst Ref as mentioned and realized scalable product/technology output. Air electrodes of MAFCs comprised at least three layers with total thickness ~3 mm, including catalytic membrane, water-proof membrane and copper-mesh collector layer.

## 3. Results and discussion

### 3.1 Morphology and structures

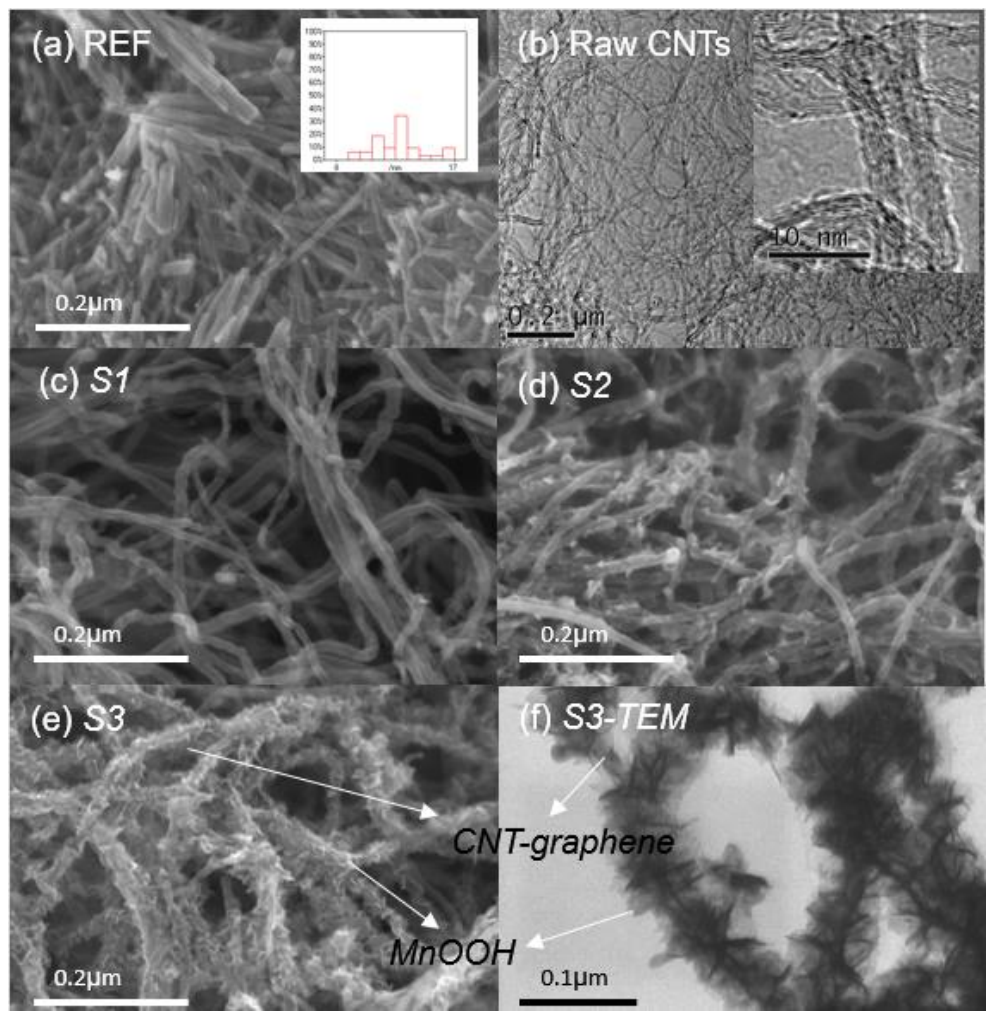

**Fig. 1. (a) manganese oxide from MITBJ for reference and its diameter distribution (inset), (b) raw CNTs under TEM with locally high-magnified image (inset), (c)~(e) a series of CMnCs with precursor KMnO$_4$ of different amounts: 0.1 mol/l for S1, 0.2 mol/l for S2 and 0.4 mol/l for S3, (f) an image of S3 under TEM.**

The CMnCs and references (Raw CNTs, Ref) were investigated using SEM (Fig. 1). The Ref displayed nano rods/wires with diameters at ~12.7 nm and lengths up to several micrometers (<1.5 μm). Large agglomerates/bundles for Ref were found (Fig. 1a) [6]. It's understandable that strong nano-size effect among these rods/wires existed, and reduced the specific active surfaces as well as consequent oxygen adspecies for enhancing ORR reactions. While the CNTs were over 10 μm in length and ~10 nm in diameter (Fig. 1b), considerably longer and finer than the Ref. At a high magnification under TEM, CNTs were clearly observed with multiple graphene walls, coincident to relative research [11]. In our synthesis, excess CNTs were added in acidic solutions which had already dissolved a varying amount of permanganate salts. As a result, manganese oxide nano flakes were found in different densities attaching to walls (Fig. 1c-e), demonstrative of concentration dependence. From Fig. 1f, these flakes demonstrated wrapping inner tubes, having constructed a shell structure. As mentioned above, permanganate is capable of strongly oxidizing carbon in the acid condition. Septivalent manganese deprived electrons from the outer-wall carbon atoms and got reduced to lower-valence. In the meantime, nano-sized oxide seedlings in-situ formed on those defective sites, then crystallized large into flakes. Simultaneously, these graphene outer walls were being etched and degenerated into nano pieces. Some of them partially attaching the inert tubes became the catalyst supporters [19, 22]. And also it's easy to know that under an external force (during the make of electrodes), these flakes got worn away at the CNT- CNT junction zones, similarly to that pencil lead when hand-writing on papers, benefitting CNTs inter-lapping for construction of electron-delivering network inside electrodes.

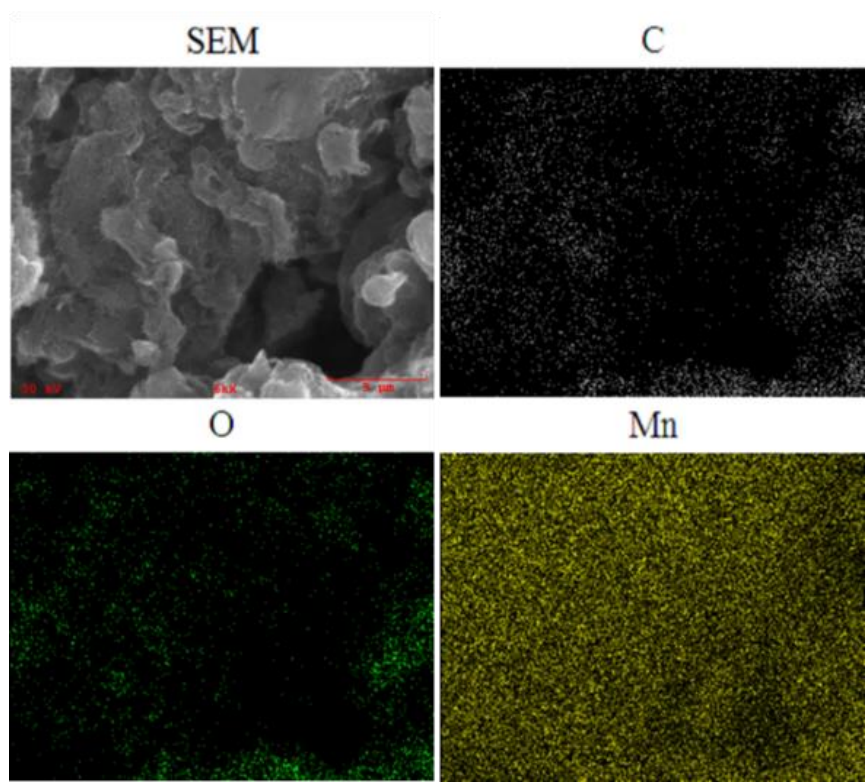

**Fig. 2. A typical image of the CMnC S3 (up-left) in magnification of ×6000 and its elemental mapping towards Carbon (up-right), Oxygen (down-left), and Manganese (down-right).**

The elemental mapping (Fig. 2) demonstrated the distribution of carbon, oxygen and manganese on the sample S3. At a low magnification (×6000) with a scale bar ~5 μm, it's hard to tell single CNT apart from the whole. Morphology-dependent distribution was observed for all elements. Together with the Fig. 1, manganese oxide material was obtained as expected with the CNTs as the supporter. Electrochemical performances

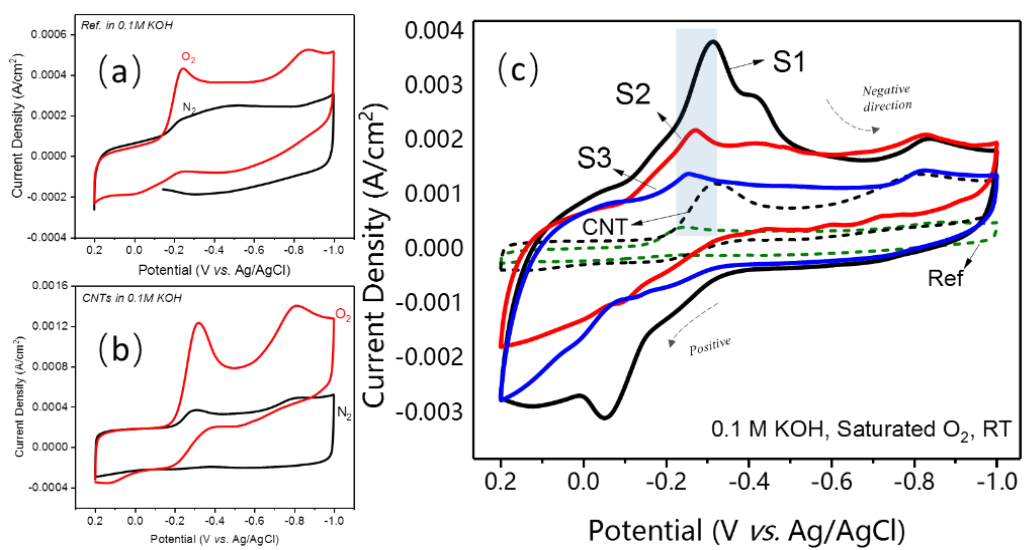

**Fig. 3. CVs of catalysts (a: Ref, b: CNTs, c: CMnCs) in $O_2$/$N_2$-saturated 0.1 M KOH solution at 50 mV/s, with a counter electrode of 20 mm×20mm platinum foil and a Ag/AgCl/saturated KCl reference electrode, temperature ~20℃. Note: the shadow area intends to highlight the onset potentials of these ORR electrodes. The data were as received from the electrochemical apparatus without further treating, the cathodic current was positive in negative direction.**

The ORR activities were laboratory tested by a three-electrode electrochemical system in a typical voltammetry mode (Fig. 3,). Catalysts were all film-coated onto glassy carbon electrodes and tested in oxygen or nitrogen saturated aqueous electrolytes in a potential window -1.0 - 0.2 V versus Ag/AgCl at 50 mV/s. The Ref and CNTs were ever both reportedly [8, 23, 24] and experimentally verified ORR sensitive (Fig. 3a, b). Defects of the CNTs promote to activate the dissociation of oxygen once they approach] the surfaces of CNTs and would acquire negative charges, like pentagons at CNTs tips, pentagon-heptagon defect pairs, and even curvature in CNTs [23]. As a result, the activity of CNTs was almost twice higher than the Ref (CNTs, ~1.2 mA/cm$^{-2}$ *vs.* Ref, ~0.4 mA/cm$^2$ at around -0.3 V). The raw CNTs exhibited competent in the ORR use.

Distinct successive cathodic peaks existed if oxygen-saturated electrolyte was used for the Ref (-0.246 and -0.871 V) and CNTs ( -0.320 and -0.820 V). These were in agreement with previous reports where GC was used as catalyst supporters and four-electron reduction mechanism dominated the reactions.[5, 24-26] In detail, the peaks at around -0.3 V implied the electrochemical mediation of molecular oxygen by the oxygen-containing groups (e.g., quinone-like groups) on the GC, with superoxide ($O_2^{\bullet-}$) as the intermediate, while the sequent peaks at around -0.8 V related to a direct two-electron reduction evolution to $HO_2^-$ (superoxide also). These intermediates were followed disproportionated into $OH^-$ by the MnOOH or CNTs [24, 26].

Notably, it was the bad electrical conductivity that restrained the Ref. The proposed CMnCs featured the nano structure of oxide flakes as the "branches" with CNTs network as the "backbones". Thus the inner contact resistance was decreased and consequent electrocatalytic performance got enhanced [2]. As-obtained CMnCs performed better than the CNTs. And all I-V behaviors (Fig. 3c) of them were found alike to the references, dominated with a four-electron oxygen reduction process.

We found at different K/C ratios, the first cathodic peaks of all CMnCs (the shadow zone) respectively appeared at -0.314 V (S1), -0.273 V (S2), to -0.255 V (S3), having the tendency to shift to -0.246 V (Ref). Such peaks were speculated to comprise the simultaneous reductions of manganese and $O_2$: the transition of $Mn^{3+}$ to $Mn^{2+}$ and the formation of peroxide anions from $O_2$ [2, 4]. And the reduction degree of manganese closely related to electron transporting through CNTs network. In contrary to the S1 and S2, the CNTs of S3 suffered heavier oxidization, and output current was consequently effected the lowest. Well design of network supporter for catalyst hosts showed of great importance. Additionally, benefitting from more accessible sites (SSA), better conductive networks and involvement of graphene nano pieces, the CMnC S1 produced superior reduction/oxidation power to the others. An obvious shoulder at the first cathodic peak for the S1 to demonstrated the contribution assigned to those graphene nano pieces, which was around -0.4 V [7]. A further intercalation of CNTs at higher K/C ratios, however in our mind, lessened the role of graphene (a further work being undergoing). Therefore, the peak current correspondingly decreased obviously.

### 3.2 Power generation of CMnCs AFC

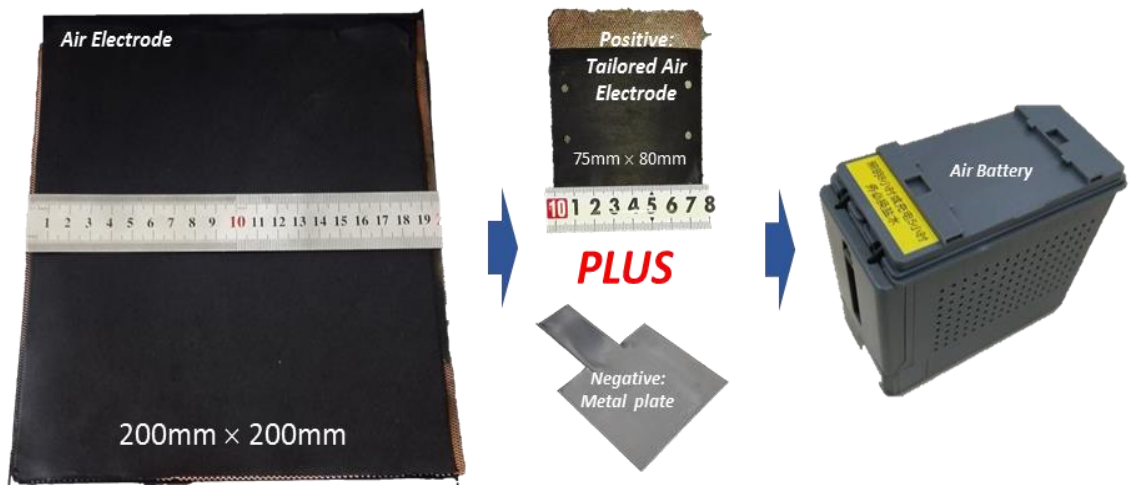

**Fig. 4. Assembly of AFC: (left) large factory-made air electrode with sides at 200 mm long, thickness at ~3 mm; (mid-top) a tailored compatible sheet electrode in a smaller size 75mm × 80mm, configured with a commercial metal (e.g., zinc) plate as negative electrode (mid-bottom); (right) an assembled air fuel cell product.**

Assembly of one single cell was briefly unveiled (Fig. 4). With self-designed automatic continuous production line, the factory is able to satisfy users to custom fabricate electrode sheets with requested sizes. As shown, the air electrodes were uniformly 200 mm× 200 mm. As anodes they were tailored into ~75 mm× 80 mm. The CMnCs based AFC were obtained. Then typical tests in factory were done, one of which adopted the Zn-air system with 6 M KOH as aqueous electrolyte (Fig. 5) [27]. In Fig. 8a, the discharge voltage and areal output power were plotted as a function of the current density. In general, all discharge voltages gradually went downward as the applied current densities increased. Comparably, the S1 AFC discharged at a current ~152.8 mA/cm$^{-2}$ at 1 V (the factory benchmark), the highest over the others. The S2, S3 and Ref AFCs performed rather close to each other (~121 mA/cm$^{-2}$). Meanwhile, the voltage drop rate of the S1 demonstrated the smallest, yet with a significant polarization throughout the whole discharge. Unexpected, the other CMnCs AFCs were even worse than the Ref.

Discharge power investigation were subsequently followed and the inflection points appeared at over 180 mA/cm$^2$. The S1 AFC output a maximal power ~174.83 mW/cm$^{-2}$, higher than the others. Interestingly, the S3 generated more power than the S2 at higher current density over 100 mA/cm$^2$. This distinction was ascribed to current recipe of catalytic membrane, which involved a majority of highly conductive substances. They not only helped decrease the resistance, but also made full use of these active sites through physically contacting with oxide nano flakes for S3, since the oxide flakes took more proportion over the S2 from TG analysis (not shown).

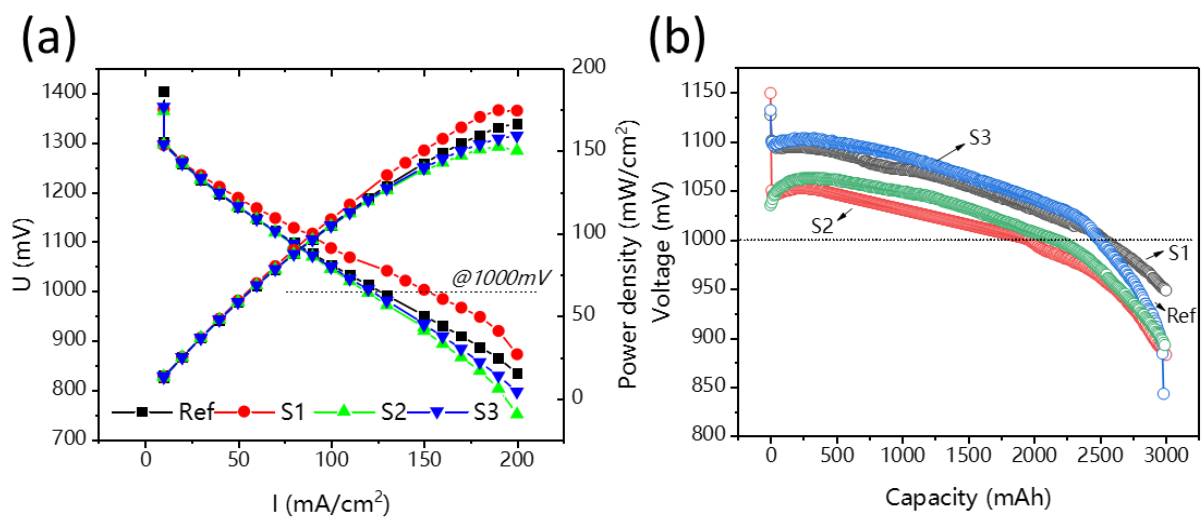

**Fig. 5. (a) Output voltages and power density as a function of current densities, (b) the 3$^{rd}$ galvanostatic discharge with areal output current density at 100 mA/cm$^2$.**

We cycled these cells through galvanostatic discharging several times at 100 mA/cm$^{-2}$. To start new discharge, anode plate and electrolyte would be renewed to ensure an identical condition. Each operation was terminated at the discharge capacity reaching 3000 mAh (~ 3 hours). Used cathodes were found performance-enhanced in next cycle. The reason was speculated linked to the electrolyte which fully penetrated inside electrodes to lower the cell resistance. This was verified by the voltage increase (for Ref, from 1.054 V to 1.062 V; S1, 1.087 V to 1.094 V; S2, 1.045 V to 1.052 V; S3, 1.050 V to 1.103 V). After two cycles, the S3 AFC turned to exceeding all, although the

S1 AFC remained superior (Fig. 5b). However, a drastic voltage drop was found for the S3 in the tail, even beneath the Ref. We attributed this finding to the pore pathway blockage or deformation by many insoluble resultants, since nano/micro structures were the majority for the S3 (Not shown). To verify this point, relative work is currently undergoing in the joint lab.

## 4. Conclusion

Commercial CNTs were one dimensional, found structurally defective. A mild redox reaction using $KMnO_4$ was proposed applied to treat CNTs at room condition. CNTs were easily oxidized. And some graphene nano pieces emerged on outer walls of CNTs and partially attached to the inner tubes so as to form CNTs-Graphene complex. During such, manganese oxide also formed and in-situ grew into flakes on the surfaces of these tubes/graphene plates. As a consequence, the ORR-oriented CNTs (Graphene)-MnOx composite catalysts were obtained, namely CMnCs. Factory-made AFC incorporated with these CMnCs air cathodes demonstrated to outperform targeted current commercial fuel cells. This research also intended to put weight on the importance of smart air electrode structural design.

## Acknowledgements

The authors sincerely acknowledge the financial support from Beijing University of Technology (No. 105000514116002, 105000546317502) and pilot product assembly & test supports through the advanced energy materials joint project with Materials and Industrial Technology Research Institute Beijing.